\documentclass[12pt,nofootinbib]{revtex4}
\usepackage{amssymb}
\usepackage{amsmath, graphicx}
\usepackage{slashed}
\usepackage[unicode]{hyperref}
\usepackage{csquotes}
\usepackage{braket}
\usepackage{natbib}
\usepackage{array}
\usepackage{xcolor}
\usepackage{multirow}
\usepackage{pifont}
\newcommand{\xmark}{\ding{55}}
\newcommand{\cmark}{\ding{51}}

\begin{document}
\title{Principle of no paradoxical Universe}
\author{Ngo Phuc Duc Loc}
\thanks{Email: locngo148@gmail.com}

\affiliation{Physics Division, National Center for Theoretical Sciences,
National Taiwan University, Taipei 106319, Taiwan}

\begin{abstract}
In the standard $\Lambda\rm CDM$ model, the Universe will keep expanding forever and its lifetime is infinite. Therefore, even the heaviest supermassive black hole will eventually evaporate and the universal occurrence of the black hole information paradox is unavoidable. However, the recent DESI results seem to indicate that dark energy is dynamical and, more importantly, it is possible that the Universe has a finite lifetime and will experience a ``Big Crunch" in some billion years from now. In light of this, the lightest confirmed black hole with mass of order one solar mass may not have a chance to evaporate before the Universe ends. Inspired by these observations, we thus raise an intriguing conjecture: Does the Universe choose a configuration such that the ``in-principle" black hole information paradox can never actually happen in reality? We derive a lower bound on black hole mass as a function of the Universe's lifetime such that the information paradox cannot be realized in nature and some resulting predictions are made. Some profound implications for other related big ideas in physics are briefly discussed.
\\

\end{abstract}

\maketitle

\tableofcontents

\section{Introduction}

At the end of the 20th century, observations of type Ia supernovae indicated that the expansion of the Universe is accelerating, and that the acceleration can be described by a cosmological constant called $\Lambda$ \cite{SupernovaCosmologyProject:1998vns,SupernovaSearchTeam:1998fmf}. Generally speaking, we use the term dark energy (DE) to indicate any substance that could be responsible for late-time acceleration, and the cosmological constant is one of the candidates\footnote{Unlike the usual matter and energy, the energy density of the cosmological constant does not dilute with the expansion of space as if it is an intrinsic feature of space itself, and thus vacuum energy of quantum fields appears to be the most natural physical origin of this cosmological constant except for the fact that the predicted value is off by $\sim 120$ orders of magnitude compared with observation! Therefore, even if DE is a cosmological constant, we need a radical explanation for this.}. Together with the observations of cold dark matter (DM) via various measurements \cite{Cirelli:2024ssz}, we arrived at the so-called $\Lambda \rm CDM$ model of cosmology. This model has passed through numerous measurements and is now widely regarded as the most successful model of the Universe at large scales. However, the recent observation of baryonic acoustic oscillation via the Dark Energy Spectroscopic Instrument (DESI) has revealed that DE seems to be evolving \cite{DESI:2025zgx}. This renewed tremendous interests in some decades-old ideas of dynamical DE.

But DE is not the only mysterious gravitational aspect of the Universe. Black holes (BH), which could also be a DM candidate in certain mass range, offer a natural laboratory where we can check the consistency of known physical laws. The extreme geometry of the BH inevitably leads to particle productions via Hawking radiation \cite{Hawking:1974rv,Hawking:1975vcx}. In the standard picture, BH eventually evaporates completely and this will lead to the infamous BH information paradox where the unitary principle of quantum mechanics cannot be simultaneously consistent with the equivalence principle of relativity \cite{Almheiri:2012rt,Almheiri:2020cfm,Hawking:1976ra}. It is important to realize, however, that this inconsistency only happens after a given timescale known as the ``Page time" which is the time at which BH has lost half of its initial degrees of freedom.

In the standard $\Lambda\rm CDM$ model of cosmology, the Universe is currently, and will forever be, dominated by DE that is constantly ``created" by the expansion of space itself. Thus, the Universe will eventually end up in a ``Big Rip" scenario where things are separated from each other at a faster and faster rate. The lifetime of the Universe is infinite, or perhaps at least until the definition of time to describe changes breaks down when everything loses all of its energy in a heat death. In this scenario, therefore, we don't have to question whether or not the BH information paradox will ever happen because it's just a matter of time. Even the heaviest supermassive BH will eventually evaporate and the universal occurence of BH information paradox across the Universe is unavoidable. However, the DESI results allow for a ``Big Crunch" scenario in which the Universe could recollapse to a singular point after some amount of time. The Universe's lifetime in this scenario is finite and we are curious if the BH information paradox would ever happen in the real world.

The rest of this paper is organized as folllows. We review the BH information paradox and estimate the Page time in Sec. \ref{sec: BH information paradox}. We discuss some simple benchmark models of dynamical DE to estimate the lifetime of the Universe in Sec. \ref{sec: lifetime of Universe} with details given in Appendix \ref{sec:basics of dynamical DE}. The statement of the ``principle of no paradoxical Universe" is given in Sec. \ref{sec: principle of no paradoxical Universe} and some resulting predictions are presented in Sec. \ref{sec:predictions}. We summarize and provide some interesting final thoughts in Sec. \ref{sec:discussion}. A proposed lower bound on population to claim existence of BH is given in Appendix \ref{appendix: beta_min}. We use the natural units in which $\hbar=c=k_B=1$.

\section{Black hole information paradox}\label{sec: BH information paradox}
The origin of BH information paradox lies at the inconsistency between the perspectives of infalling and faraway observers, which nonetheless only happens after a given timescale known as the ``Page time". While there are different ways to understand this paradox, let us focus on the simplest intuitive version. Consider BH described by a pure quantum state. The BH has an event horizon. Inside the horizon, not even light can escape. Outside the horizon, things can still escape to infinity. The event horizon itself is a smooth boundary and the infalling observer would not notice anything dramatic when crossing through it. The initial Bekenstein-Hawking entropy of BH is given by \cite{Bekenstein:1973ur,Hawking:1975vcx}
\begin{equation}
    S_\text{BH,i}=\frac{A_i}{4G}=\frac{\pi R_\text{Sch,i}^2}{G}=4\pi GM_i^2,
\end{equation}
where $A_i$ is the initial area of BH event horizon, $G$ is the gravitational constant, $R_\text{Sch,i}$ is the initial Schwarzschild radius, and $M_i$ is the initial BH mass. This entropy describes the number of available degrees of freedom (d.o.f.) inside the BH. In particular, $e^{S_\text{BH}}$ is the number of degenerate microstates that describe the same macrostate of BH with mass $M$. We have two equally true statements:
\begin{enumerate}
\item In quantum field theory, it is known that spatially nearby d.o.f. are entangled. From the perspective of the observer near the event horizon, as the BH evaporates, the outgoing Hawking quanta must be entangled with the infalling quanta as they were closed to each other at the event horizon. A smooth event horizon implies that there is no ``firewall" to prevent entanglement generation \footnote{Alternatively, by invoking the equivalence principle of general relativity, one sees that the effects experienced in a gravitational field are locally indistinguishable from those in a uniformly accelerated frame. Because a uniformly accelerated frame does not spontaneously generate a ``firewall", the equivalence principle implies that a black hole event horizon cannot do so either.}.
\item As seen from the outside faraway observer, BH is nothing but an ordinary quantum system that must evolve unitarily under the laws of quantum mechanics. Namely, the initial pure quantum state of BH must remain pure. By the demand of unitary evolution of BH, the Hawking radiation and the BH's remaining interior d.o.f. are both in the mixed states due to their mutual entanglement, but the total state as a whole must remain pure.
\end{enumerate}
The above two perspectives are perfectly consistent as long as the ``age" of BH is less than the Page time. The age of BH is defined as the amount of time elapsed since the BH formation, as measured by the faraway observer. The Page time, also as measured by a faraway observer, is defined as the time at which BH has lost half of its initial d.o.f. \footnote{Recall that the dimensionality of the total Hilbert space of two subsystems is the product of subsystems' dimensionality, so the number of d.o.f. of half of the total system is $e^{S_\text{BH}/2}$. Also, there is another version of the Page time which is instead defined as the moment when the entanglement entropy of BH equals its Bekenstein-Hawking entropy \cite{Page:2013dx,Perez-Gonzalez:2025try}. This definition only slightly affects the value of the Page time and should not concern us here.}: $S_\text{BH,i}\rightarrow S_\text{BH,i}/2$ so that $M_i\rightarrow M_i/\sqrt{2}$. For a ``young" BH whose age is less than the Page time, Hawking radiation is entangled with BH's interior d.o.f. and the total state remains pure. Therefore, both statements 1 and 2 above are correct and there is no information paradox. However, for an ``old" BH whose age is greater than the Page time, the remaining interior d.o.f. were already maximally entangled with early Hawking radiation and therefore cannot be simultaneously entangled further with any late Hawking radiation. In other words, in order to preserve unitarity (statement 2), the early and the late Hawking quanta must be entangled, which nevertheless violates statement 1 as the late Hawking quanta must be entangled with BH's remaining interior d.o.f. instead. This is the essence of BH information paradox \cite{Almheiri:2012rt,Almheiri:2020cfm}.

To estimate the Page time, we write down the mass loss rate of BH \cite{Hawking:1974rv,Hawking:1975vcx,Gibbons:1977mu,Hooper:2019gtx}:
\begin{equation}
    \frac{dM}{dt}=-\mathcal{G}\frac{g_{*\text{,H}}(T_\text{H})}{30720\pi}\frac{m_\text{pl}^4}{M^2},
\end{equation}
where $\mathcal{G}\approx 3.8$ is the graybody factor \footnote{Sometimes also known as the Page factor.}, $m_\text{pl}\approx 1.22 \times 10^{19}\ \rm GeV$ is the Planck mass, $M$ here is the instantaneous BH mass at time $t$, $g_{*,\rm H}$ is the number of effective relativistic d.o.f. at Hawking temperature $T_\text{H}$ given by
\begin{equation}
    \left(\frac{T_\text{H}}{\rm GeV}\right)= 1.06\times 10^{13}\left(\frac{\rm g}{M}\right).
\end{equation}
The value of $g_{*,\rm H}$ is determined according to the following prescription \cite{MacGibbon:1990zk,MacGibbon:1991tj}:
\begin{equation}
    g_{*,\rm H}(T_\text{H})=\sum_i w_ig_{\rm i,H}
\end{equation}
where $w_i=2s_i+1$ for massive particles of spin $s_i$ and $w_i=1\ (2)$ for massless particles of spin $s_i=0\ (s_i>0)$, and
\begin{equation}
    g_{\rm i,H}=\begin{cases}
        1.82\hspace{2cm}s=0\\
        1\hspace{2.5cm}s=1/2\\
        0.41\hspace{2cm}s=1\\
        0.05\hspace{2cm}s=2
    \end{cases}
\end{equation}
As we will see later, the mass range of interest is of order $10^{15}\ \rm g$, which implies Hawking temperature of order 10 MeV. Therefore, we expect that BH can produce photons $\gamma$, neutrinos $\nu$, and electron-positron $e^\pm$ so that $g_{*,\rm H}=2\times 0.41+3\times 2\times 1+2\times 2\times 1= 10.82$. The mass loss rate is then
\begin{equation}
    \frac{dM}{dt}\simeq -4.26\times 10^{-4}\frac{m_\text{pl}^4}{M^2}.
\end{equation}
Thus, the Page time at which BH mass reduces to $M_i/\sqrt{2}$ is
\begin{equation}
     t_\text{Page}\simeq 505.83\frac{M_i^3}{m_\text{pl}^4},
\end{equation}
or in the conventional normalized form as
\begin{equation}\label{eq:Page time}
    \left(\frac{t_\text{Page}}{\rm Gyr}\right)\simeq 8.38\times 10^{-44}\left(\frac{M_i}{\rm g}\right)^3.
\end{equation}
The Page time is, of course, less than the lifetime of BH as the latter is defined as the moment when BH disappears completely. For example, a BH with mass $\sim 5\times 10^{14} \ \rm g$ has a lifetime equal to the age of the Universe \cite{Carr:2020gox}, which is around 13.8 billion years, but its Page time is about 10.5 billion years after the Big Bang, so the information paradox occurred around the time at the onset of dark energy domination.

\section{Lifetime of our Universe}\label{sec: lifetime of Universe}

In light of the DESI results, we basically have three scenarios for the ultimate fate of the Universe. The first scenario is that DE will somehow restore its strength in the future, perhaps through oscillating DE models \cite{Escamilla:2024fzq,Rezaei:2024vtg}, so that the Universe still keeps expanding forever and ends in a Big Rip. The lifetime of the Universe is infinite in this case \footnote{The concept of infinite time depends on how we define time. This profound issue has both interesting physical and philosophical  aspects, which nonetheless is not our concern for the present purpose.}. The second scenario is that the weakening DE will become even weaker in the future and its energy density will drop down close to zero but remain positive, if we naively take the DESI results at face value \cite{DESI:2025zgx}. In this case, non-relativistic matter will dominate the Universe again in the future and thus we have $H=2/3t$, with $H$ is the Hubble rate\footnote{We are assuming flat geometry.}. Therefore, the Universe will be asymptotically static as $H\rightarrow 0$ when $t\rightarrow\infty$. The Universe experiences neither a Big Rip nor a Big Crunch. The lifetime of the Universe is therefore still infinite in this case. The third scenario is that if DE density decreases and eventually flips its sign in the future, the Universe will recollapse in a Big Crunch and its lifetime is finite \cite{Luu:2025dax,Cline:2025sbt,Tada:2024znt}.

As a proof of concept, we consider a simple quadratic Hilltop quintessence model for the crunching scenario with the potential of the form $V=V_0+V_2\phi^2/2$ with $V_0>0$ and $V_2<0$ \cite{Cline:2025sbt,Shlivko:2025fgv}. The logic to get dynamical DE in this model is that the field begins to roll recently due to no longer being frozen by Hubble friction. The field keeps rolling down toward the negative regime of the potential, which therefore admits a crunching scenario. This potential is unbounded below and can be thought of as an effective description of a true potential that may have a global minimum. In any case, what we care about is that the DE density flips its sign to the level that can realize a Big Crunch. By using the best-fit values of the parameters of this model found in \cite{Cline:2025sbt}, we find in Appendix \ref{sec:basics of dynamical DE} the lifetime of the Universe to be $\tau_\text{Universe}\approx 35\ \rm Gyr$.

The model has $w\geq -1$ and does not cross the phantom regime \footnote{The statement of phantom crossing of the DESI collaboration relies specifically on the chosen phenomenological CPL parameterization. If one uses different EOS parameterizations for particular theoretical models, the phantom crossing may disappear.}. Also, given the fact that experimental data will be updated in the future and also to allow for flexibility of model buildings, we think it is better to express our predictions as a function of the Universe's lifetime. The aforementioned  benchmark model is simply meant to give a useful reference point for demonstration.

\section{Principle of no paradoxical Universe}\label{sec: principle of no paradoxical Universe}

At the time of writing, the lightest BH candidate has mass $\sim 2.6\ M_\odot$ and was detected via gravitational wave (GW) produced from binary coalescence event GW190814 \cite{LIGOScientific:2020zkf}, where $M_\odot\approx 2\times 10^{33}\ \rm g$ is the present solar mass \footnote{There are also other light BH candidates with a similar mass \cite{Barr:2024wwl,Song:2024tqr}}. At the other end, the heaviest confirmed BH is a quasar TON 618 with mass $\sim 40.7\times 10^9\ M_\odot$ \cite{Ge} \footnote{Initial estimate gave $\sim66\times 10^9\ M_\odot$ \cite{Shemmer:2004ph} but this was later refined to $\sim 40.7\times 10^9\ M_\odot$ in \cite{Ge}.}. Thus, from the current observational status, the Page time takes the range $10^{58}\ \rm Gyr<t_\text{Page}<10^{88}\ \rm Gyr$, which are extremely huge but finite numbers. The universal occurrence of BH information paradox will therefore eventually happen in an infinite-lifetime Universe, like in the standard $\Lambda \rm CDM$ model. That may not be so in a recollapsing Universe. The predicted lifetimes of the Universe of some benchmark models mentioned in the previous section are still much smaller than the Page time of astrophysical BHs, so the information paradox is automatically evaded. However, explosion of massive stars is not the only way to make BHs.

It is possible that BH could form in the very early Universe, bypassing the usual path of stellar evolution. Such BH is called primordial black hole (PBH) \cite{Hawking:1971ei,Carr:1974nx}. PBH could have a very wide mass range depending on the formation time and formation mechanism, ranging from microscopic Planck size with mass of order $10^{-5}\ \rm g$ to close to the supermassive size with mass of order $10^{5}M_\odot$ \cite{Carr:2020gox}. It is the possible existence of small-mass PBH, which has much shorter Page time, that makes the BH information paradox look more realistic even in a recollapsing Universe. However, the paradox will not occur whatsoever if the Universe crunches before the Page time of the smallest existing BHs.

We therefore propose a conjecture called ``principle of no paradoxical Universe", which states:
\begin{quote}
    \textit{``The Page time of the smallest black hole existing within the Universe must be greater than the Universe's lifetime."}
\end{quote}
If this principle holds, the information paradox can never be realized in nature because the collapse of the Universe destroys everything within it, including BHs. The final state is a (potentially) devastating singular point whose subsequent fate is beyond our current understanding. In this way, unitary principle of quantum mechanics and equivalence principle of general relativity can always be simultaneously consistent during the Universe's lifetime. While this conjecture was specifically formulated in the context of BH information paradox, its vision could be extended to a broader landscape that will be discussed a bit more in the discussion section.

Most importantly, our conjecture has testable predictions, which we discuss in the next section.

\section{Predictions}\label{sec:predictions}
By demanding that the Page time given in Eq. \ref{eq:Page time} is greater than the lifetime of our Universe \footnote{Precisely speaking, we should demand $t_i+t_\text{Page}\geq\tau_\text{Universe}$ with $t_i$ is the formation time of BH. However, the lower bound on BH mass that we will discuss is small and therefore is most relevant if BH has primordial origin, so the formation time is much smaller than the Page time and thus can be neglected.}, we arrive at the lower bound on initial BH mass \footnote{If there is accretion or merger, the allowed initial mass could be smaller than this lower bound as the Page time could be prolonged. These effects, however, may not always happen to all BHs and therefore this lower bound is to guarantee that the paradox won't occur in any scenarios.}
\begin{equation}\label{eq:M_BH}
    \left(\frac{M_i}{\rm g}\right)\gtrsim 2.29\times 10^{14}\left(\frac{\tau_\text{Universe}}{\rm Gyr}\right)^{1/3}.
\end{equation}
We remind the readers that the notation $\tau_\text{Universe}$ denotes the Universe's lifetime, not its current age. Imagine a Universe that lives for a trillion years, the minimum initial BH mass needed to avoid the occurrence of information paradox is $\sim 2\times 10^{15}\rm g$. If the Universe only lives for 13.8 billion years, which is its current age, the minimum BH mass to avoid the information paradox is $\sim 5.5\times 10^{14}\rm g$ which is slightly larger than the mass $\sim 5\times 10^{14}\rm g$ whose lifetime equals Universe's age. That is because we need the Page time, which is earlier than BH's lifetime, to be greater than the Universe's lifetime and thus the BH mass must be larger.

The benchmark crunching model that we discussed earlier predicts $\tau_\text{Universe}\approx 35\ \rm Gyr$, which implies $M_i\gtrsim 7.5\times 10^{14}\rm g$. The lower bound on BH mass is quite insensitive to the Universe's lifetime, so the minimum mass scale of order $\mathcal{O}(10^{15}\rm g)$ is quite robust against various models. We will therefore use this threshold value to make the following predictions.

\subsection{Particles evaporated from BHs must have energy less than 10 MeV}

The first obvious resulting consequence is that particles produced from BH's evaporation must have energy less than $\sim 0.01\ \rm GeV$. Any higher-energy cosmic rays must have come from other astrophysical sources and not BHs. The recent ultra-high-energy neutrino event with energy of order $\sim 10^8\ \rm GeV$ observed at KM3NeT has received great interests \cite{KM3NeT:2025npi}. A possible explanation for this could be that the neutrino was evaporated from PBH at the late stage of its evaporation \cite{Klipfel:2025jql}. However, this scenario was then rejected by \cite{Airoldi:2025opo} when they realize that we did not receive the expected gamma rays from such explosive PBH. This fact supports our conjecture. We do not expect BH to cross the Page time during the Universe's lifetime.

\subsection{Our conjecture dictates which GW sources can exist, and which cannot}

There are also more distinctive predictions of stochastic GW backgrounds. Our current understanding of stellar physics indicates that subsolar mass BHs cannot be produced from explosion of massive stars. Any evidence of subsolar mass BHs implies that they must have a primordial origin and are therefore PBHs. A population of PBHs distributed across the Universe creates \textit{unique} GW spectra that could be tested in the upcoming GW observatories. In particular:
\begin{itemize}
    \item \textbf{Formation}:\\
    PBHs are typically formed from enhancement of curvature perturbation at small scale, which will inevitably induce second-order tensor perturbation known as scalar-induced gravitational waves (SIGW) \cite{Kohri:2018awv}. PBHs heavier than $10^{9}\rm g$ have lifetime greater than 1 second, which is the timescale for the onset of Big Bang Nucleosynthesis (BBN) and the subsequent well-known thermal history, and therefore must be subdominant at early times. In this case, the amplitude of SIGW is not suppressed and we expect $\Omega_\text{SIGW}^\text{peak}\sim\mathcal{O}(10^{-10}-10^{-9})$. Entropy is conserved in this case, which implies the peak frequency \cite{Loc:2024qbz}
    \begin{equation}
        \frac{f_\text{SIGW}^\text{peak}}{\rm Hz}=2.53\times 10^8\gamma^{1/2}\left(\frac{\rm g}{M_i}\right)^{1/2},
    \end{equation}
    where $\gamma\simeq0.2.$ We will have some more to talk about SIGW later in subsection \ref{subsec:null test}.

    \item \textbf{Mergers}:\\
    If PBHs form binary systems, they can produce a stochastic GW background during the inspiralling  phase of PBH mergers. The GW spectrum from PBH mergers has a distinctive form $\Omega_\text{GW}^\text{merger}\propto f^{2/3}$ for $f<f_\text{peak}^\text{merger}$ and rapid drop for $f>f_\text{peak}^\text{merger}$ when the inspiralling phase ends. The peak frequency is given by \cite{Phinney:2001di}
    \begin{equation}
        \left(\frac{f_\text{peak}^\text{merger}}{\rm Hz}\right)\simeq 4.4\times 10^{36}\left(\frac{\rm g}{M_i}\right).
    \end{equation}
    This expression is valid for BHs that are still existing today, which is the case for our conjecture. If PBHs evaporated in the radiation-dominated (RD) or matter-dominated (MD) phase, the dependence of peak frequency on BH mass would be different \cite{Loc:2026dto}.

    \item \textbf{Evaporation:}\\
    If gravity can be quantized, quantum excitations of gravitational field are called ``gravitons". These spin-2 particles could be evaporated from BHs and create a GW spectrum at the coarse-grained level. There are two cases \cite{Loc:2026dto}:
    \begin{itemize}
        \item If $M_i\lesssim 5\times 10^{14}\rm g$, BH's lifetime is less than the age of the Universe and therefore they evaporated completely. In this case, the GW spectrum from gravitons scales as $\Omega_\text{GW}^\text{graviton}\propto f^{5/2}$ for $f<f_\text{peak}^\text{graviton}$ and $\Omega_\text{GW}^\text{graviton}\propto f^{-1}$ for $f>f_\text{peak}^\text{graviton}$. This is a unique spectral form of evaporated BHs originated from genuine BH physics and is independent from the cosmology. We predict null detection of this kind of GW since our conjecture does not allow for the existence of fully evaporated BHs.
        \item If $M_i\gtrsim 5\times 10^{14}\rm g$, BH's lifetime is greater than the age of the Universe and therefore they are still evaporating now. In this case, the GW spectrum from gravitons has a ``premature" form when the high-frequency tail has not yet been fully developed. That is, $\Omega_\text{GW}^\text{graviton}\propto f^{5/2}$ for $f<f_\text{peak}^\text{graviton}$ and rapid drop for $f>f_\text{peak}^\text{graviton}$. Our conjecture still allows for the existence of this kind of GW spectrum but only for BHs with mass satisfying the lower bound in Eq. \ref{eq:M_BH}. The prediction for the peak frequency of GW spectrum made of graviton is identical to that of GW spectrum from BHs merger up to an order one factor: $f_\text{peak}^\text{graviton}/f_\text{peak}^\text{merger}\approx 2.3$ \cite{Loc:2026dto}.
    \end{itemize}
    We should emphasize that the existence of the so-called ``gravitons" crucially relies on the assumption of canonical quantization of gravity. If the underlying structure of spacetime is emergent or gravity is fundamentally different from gauge interactions, there may be no such gravitons. 
    
    \item \textbf{No GW backgrounds associated with population fluctuation of evaporated ultra-light PBHs:}
    \begin{itemize}
    \item If ultra-light PBHs with mass below $10^9\rm g$ temporarily dominated the early Universe before BBN, their rapid decay would convert initial isocurvature perturbation between PBHs and radiation into curvature perturbation that would induce second-order GW background \cite{Papanikolaou:2020qtd,Domenech:2020ssp,Loc:2025whx}. This GW source takes the form $\Omega_\text{GW}^\text{isocurvature}\propto f^{11/3}$ for $f<f_\text{UV}$, where the cutoff frequency $f_\text{UV}$ is determined by PBHs' mean separation. We predict null detection of such signal because no fully evaporated BHs should exist according to our conjecture.
\item A period of PBH domination also induces another kind of GW background from adiabatic perturbation via the so-called poltergeist mechanism \cite{Inomata:2020lmk,Inomata:2025wiv}. GW is effectively produced due to the sudden change of gravitational potential from large constant value during PBH domination to rapid oscillation during subsequent RD phase when PBHs evaporate and reheat the Universe. The spectrum features a sharp peak at $f_\text{max}=\min\{f_\text{dom},f_\text{NL}\}$ with $f_\text{dom}$ is the mode entering horizon at the onset of PBH domination and $f_\text{NL}$ is the mode that would become nonlinear by the end of such period \cite{Inomata:2019ivs,Allahverdi:2025btx}. We predict null detection of such signal because no fully evaporated BHs should exist according to our conjecture.
    \end{itemize}
\end{itemize}

\begin{table}[h!]
    \centering
    \begin{tabular}{|c|c|c|c|}
    \hline
       \multicolumn{2}{|c|}{\textbf{GW sources}}  & \textbf{Existence} & \textbf{Notable features} \\
       \hline
        \multicolumn{2}{|c|}{Formation} & \cmark & $
        \begin{aligned}
                    &\text{If SIGW, typically symmetric peak,}\\
                    &\Omega_\text{SIGW,peak}\sim \mathcal{O}(10^{-10}-10^{-9})
        \end{aligned}$\\
        \hline
        \multicolumn{2}{|c|}{Merger} & \cmark & $
        \begin{aligned}
            &\Omega_\text{GW}^\text{merger}\propto f^{2/3}\ (f<f_\text{peak}^\text{merger})\\
            &\text{rapid drop for }f>f_\text{peak}^\text{merger}
        \end{aligned}
        $\\
        \hline
\multirow{2}{*}{
$
\begin{aligned}
    &\text{ Evaporation}\\
    &\text{(questionable existence}\\
    &\text{ of spin-2 gravitons)}
\end{aligned}
$} & 
$
\begin{aligned}
    &\text{Full spectrum}\\
    &(M_i\lesssim 5\times 10^{14}\rm g)
\end{aligned}
$
& \xmark & $
\begin{aligned}
    &\Omega_\text{GW}^\text{graviton}\propto f^{5/2} \ (f<f_\text{peak}^\text{graviton})\\ &\Omega_\text{GW}^\text{graviton}\propto f^{-1}\ (f>f_\text{peak}^\text{graviton})
    \end{aligned}
$ \\
\cline{2-4}
& 
$
\begin{aligned}
    &\text{Premature spectrum}\\
    &(M_i\gtrsim 5\times 10^{14}\rm g)
\end{aligned}
$
& \cmark & $
\begin{aligned}
&\Omega_\text{GW}^\text{graviton}\propto f^{5/2}\ (f<f_\text{peak}^\text{graviton})\\
&\text{rapid drop for }f>f_\text{peak}^\text{graviton}    
\end{aligned}
$\\
\hline
\multicolumn{2}{|c|}{PBH-induced isocurvature} & \xmark & $\Omega_\text{GW}^\text{isocurvature}\propto f^{11/3}$\\
\hline
\multicolumn{2}{|c|}{PBH-induced sudden transition} & \xmark & Sharp peak at $f_\text{max}=\min\{f_\text{dom},f_\text{NL}\}$\\
\hline
\end{tabular}
    \caption{Summary of GW predictions of our conjecture. For GW sources that could exist, the peak frequencies must satisfy bounds in Fig. \ref{fig:predictions}.}
    \label{tab:summary}
\end{table}

\begin{figure}[h!]
    \centering
    \includegraphics[width=0.51\linewidth]{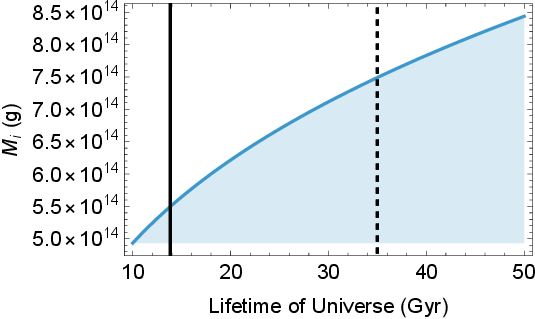}
    \includegraphics[width=0.47\linewidth]{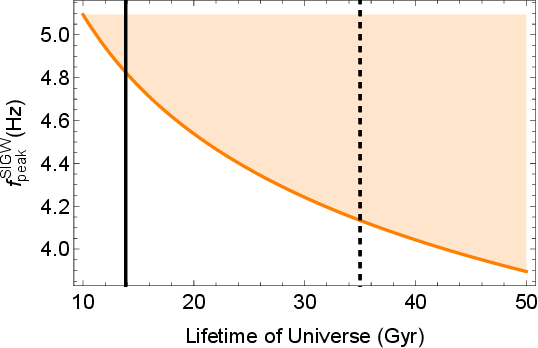}
    \includegraphics[width=0.48\linewidth]{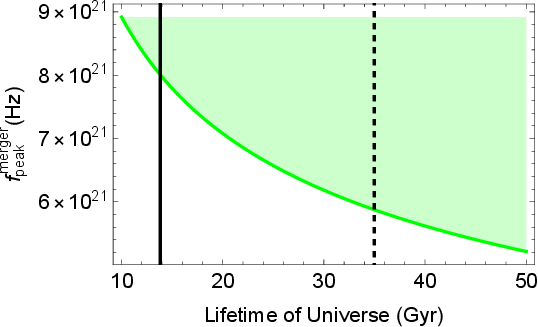}
    \includegraphics[width=0.5\linewidth]{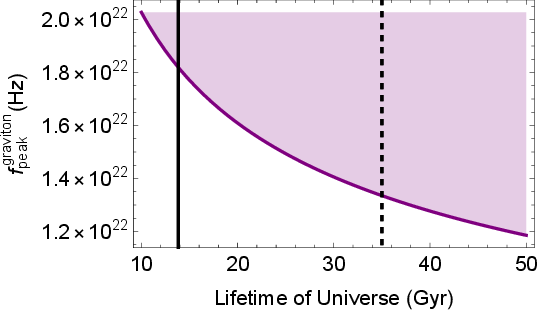}
    \caption{Observational predictions of the principle of no paradoxical Universe. The shaded regions are excluded. The dashed black line corresponds to $\tau_\text{Universe}\approx 35\ \rm Gyr$ of the quadratic Hilltop quintessence model given in Appendix \ref{sec:basics of dynamical DE} of this paper. The solid black line is the current age of the Universe $t_\text{age}\approx 13.84\ \rm Gyr$, which is obviously its minimum lifetime.}
    \label{fig:predictions}
\end{figure}

We summarize in Table \ref{tab:summary} the GW-related observational predictions of our conjecture. It should be noted that, under typical conditions, the GW spectra feature a universal $\Omega_\text{GW}\propto f^3$ scaling at the low-frequency (large-scale) regime \cite{Cai:2019cdl}, but the regime around the peak captures the distinctive physics of the GW sources and is detectable. For GW sources that could be detected, the peak frequencies must satisfy bounds in Fig. \ref{fig:predictions}. The shaded regions are excluded. The solid black line is the current age of the Universe, which is also its minimum lifetime. The dashed black line corresponds to $\tau_\text{Universe}=35\ \rm Gyr$ predicted by the quadratic Hilltop quintessence model discussed in Sec. \ref{sec: lifetime of Universe}.

\subsection{A null test to confirm our conjecture}\label{subsec:null test}

To confirm our proposed conjecture, null detection of SIGW associated with formation of PBH with mass violating the bound in Eq. \ref{eq:M_BH} is the most important one. BHs with mass below $\sim 10^{15}\rm g$, if exist, must have a primordial origin. Therefore, we first need to recall some basic frameworks of PBH physics. PBHs formed in a standard RD phase before BBN have the initial mass \cite{Carr:2020gox,Loc:2025mzc}
\begin{equation}
    M_i=\gamma M_\text{H,i}=\gamma \frac{1}{2GH_i}=\gamma\frac{\sqrt{45}m_\text{pl}^3}{4\pi^{3/2}g_{\rho,i}^{1/2}T_i^2}
\end{equation}
\begin{equation}\label{eq:Mi vs Ti}
    \Rightarrow\frac{M_i}{\rm g}=9.45\times 10^{31}\gamma\left(\frac{106.75}{g_{\rho,i}}\right)^{1/2}\left(\frac{\rm GeV}{T_i}\right)^2,
\end{equation}
where $M_\text{H,i}$ is the horizon mass at the formation time, $\gamma\simeq 0.2$, $m_\text{pl}=1.22\times 10^{19}\rm GeV=2.18\times 10^{-5}\rm g $ is the Planck mass, $g_{\rho,i}$ is the number of relativistic d.o.f. in energy density at the formation temperature $T_i$ and we have $g_{\rho,i}=106.75$ for PBH mass below $10^{15}\rm g$ assuming Standard Model's number of species. The PBH fraction at the formation time is defined as $\beta\equiv\rho_\text{PBH,i}/\rho_\text{rad,i}$, where $\rho_\text{PBH,i}$ is initial energy density of PBH and $\rho_\text{rad,i}$ is initial energy density of radiation. $\beta$ is related to the amplitude of primordial curvature perturbation via the relation \cite{Allahverdi:2020bys,Harada:2017fjm}
\begin{equation}
    \beta\simeq\text{Erfc}\left[\frac{0.67}{P_\zeta^{1/2}}\right],
\end{equation}
where Erfc is the complementary error function and $P_\zeta$ is the power spectrum of curvature perturbation. We proposed in Appendix \ref{appendix: beta_min} a novel lower bound on PBH fraction to claim existence of PBH 
\begin{equation}
    \beta_\text{min}\simeq 8\times 10^{-83}\gamma^{-1/2}\left(\frac{M_i}{\rm g}\right)^{3/2}.
\end{equation}
PBH with population below this limit may exist outside of the current observable Universe, but that cannot be confirmed by any means and thus we can conclude that they do not exist in nature.

Since PBH dilutes slower than radiation ($\rho_\text{PBH}\propto a^{-3}$ vs $\rho_\text{rad}\propto a^{-4}$ with $a$ is the scale factor), there exists a critical value of PBH fraction above which there could be a period of PBH domination \cite{Loc:2025mzc}
\begin{equation}\label{eq:betac}
    \beta_c=1.78\times 10^{-6}\gamma^{-1/2}\left(\frac{\rm g}{M_i}\right).
\end{equation}
Let $T_\text{dom}$ be the cosmic temperature at which PBH begins to dominate, and $T_\text{eva}$ be the cosmic temperature at which PBH evaporates. We have some well-known useful relations \cite{Loc:2025mzc,Loc:2025whx}
\begin{equation}\label{eq:beta vs temp ratio}
    \beta=\frac{T_\text{dom}}{T_i},
\end{equation}
\begin{equation}\label{eq:betac vs temp ratio}
    \beta_c=\frac{T_\text{eva}}{T_i},
\end{equation}
\begin{equation}\label{eq:Teva}
    \frac{T_\text{eva}}{\rm GeV}=1.73\times 10^{10}\left(\frac{\rm g}{M_i}\right)^{3/2}.
\end{equation}
As we mentioned in the previous subsection, a period of PBH domination would induce isocurvature GW with a strength that may violate $\Delta N_\text{eff}$ constraint on extra relativistic d.o.f. . Thus, there exists an upper bound on PBH fraction \cite{Domenech:2020ssp,Loc:2025whx}
\begin{equation}
    \beta_\text{max}=4.42\times 10^{-4}\gamma^{-1/2}\left(\frac{\rm g}{M_i}\right)^{17/24}.
\end{equation}

We have two cases:
\begin{itemize}
    \item If $\beta<\beta_c$, there is no PBH domination. The entropy conservation argument then leads to the peak wavenumber of SIGW \cite{Loc:2024qbz}
    \begin{equation}
        \frac{k_p}{\rm Mpc^{-1}}=1.63\times 10^{23}\gamma^{1/2}\left(\frac{\rm g}{M_i}\right)^{1/2}.
    \end{equation}
    Using the conversion relation $(f/\text{Hz})=1.55\times 10^{-15}(k/\rm Mpc^{-1})$, we get
    \begin{equation}
        \frac{f_\text{SIGW}^\text{peak}}{\rm Hz}=2.53\times 10^8\gamma^{1/2}\left(\frac{\rm g}{M_i}\right)^{1/2}.
    \end{equation}
    Physically, the peak frequency of SIGW is expected to be inversely proportional to the PBH mass because larger-mass PBH forms at a later time in cosmic history, so the associated horizon size is larger and thus the induced GW has lower frequency. The peak amplitude of the SIGW spectrum is \cite{Cirelli:2024ssz,Kohri:2018awv}
\begin{equation}
    \Omega_\text{SIGW}^\text{peak}\simeq 10^{-5} P_\zeta^2
\end{equation}
This gives the correct order-of-magnitude estimate for the simplest case of a monochromatic or a narrowly peaked PBH mass function. 
\item If $\beta>\beta_c$, there is PBH domination. The peak wavenumber is redshifted as \cite{Allahverdi:2025btx}
\begin{align}
    \frac{k_p}{\rm Mpc^{-1}}&=1.59\times 10^{39}\gamma\left(\frac{T_\text{dom}}{T_i}\right)\left(\frac{T_\text{eva}}{T_\text{dom}}\right)^{4/3}\left(\frac{\rm GeV}{T_\text{eva}}\right)\left(\frac{\rm g}{M_i}\right)\\
    &=1.98\times 10^{21}\gamma^{1/3}\beta^{-1/3}\left(\frac{\rm g}{M_i}\right)^{5/6},
\end{align}
where we used Eqs. \ref{eq:betac}, \ref{eq:beta vs temp ratio}, \ref{eq:betac vs temp ratio}, \ref{eq:Teva}. In the convenient frequency language, it is
\begin{equation}
    \frac{f_\text{SIGW}^\text{peak, dom}}{\rm Hz}=3.06\times 10^6\gamma^{1/3}\beta^{-1/3}\left(\frac{\rm g}{M_i}\right)^{5/6}.
\end{equation}
The extra superscript ``dom" was added to indicate this is the case with PBH domination. Physically, we see that the frequency is shifted toward lower frequency because the Universe expands faster in a matter dominated Universe (PBH domination in our case) $a\propto t^{2/3}$ compared to a radiation dominated Universe $a\propto t^{1/2}$, and thus GW's wavelength gets stretched more which leads to lower frequency. The duration of PBH domination depends on both PBH lifetime (dictated by $M_i$) and relative fraction with radiation background (dictated by $\beta$). These factors thus modify the peak frequency from $f_\text{SIGW}^\text{peak}\propto M_i^{-1/2}$ to $f_\text{SIGW}^\text{peak, dom} \propto\beta^{-1/3}M_i^{-5/6}$. In addition, since GW propagates like radiation $\rho_\text{GW}\propto a^{-4}$, the amplitude of SIGW is suppressed by a factor of $(a_\text{dom}/a_\text{eva})$:
\begin{equation}
\begin{aligned}
    \Omega_\text{SIGW}^\text{peak, dom}&=\Omega_\text{SIGW}^\text{peak}\left(\frac{a_\text{dom}}{a_\text{eva}}\right)\\
    &=\Omega_\text{SIGW}^\text{peak}\left(\frac{T_\text{eva}}{T_\text{dom}}\right)^{4/3}\\
&=\Omega_\text{SIGW}^\text{peak}\beta^{-4/3}\beta_c^{4/3}\\
&=\Omega_\text{SIGW}^\text{peak}\ 2.16\times 10^{-8}\gamma^{-2/3}\beta^{-4/3}\left(\frac{\rm g}{M_i}\right)^{4/3},
\end{aligned}
\end{equation}
where we used $a_\text{dom}/a_\text{eva}=(T_\text{eva}/T_\text{dom})^{4/3}$ \cite{Allahverdi:2025btx} and Eqs. \ref{eq:beta vs temp ratio}, \ref{eq:betac vs temp ratio}, \ref{eq:betac}. Again we see that a prolonged duration of PBH domination, characterized by large $\beta$ and/or $M_i$, will suppress the GW amplitude more. Thus, the general tendency of SIGW associated with PBH formation with a period of PBH domination is that it will be suppressed in amplitude and be shifted toward lower frequency.
\end{itemize}

\begin{figure}[h!]
    \centering
    \includegraphics[width=\linewidth]{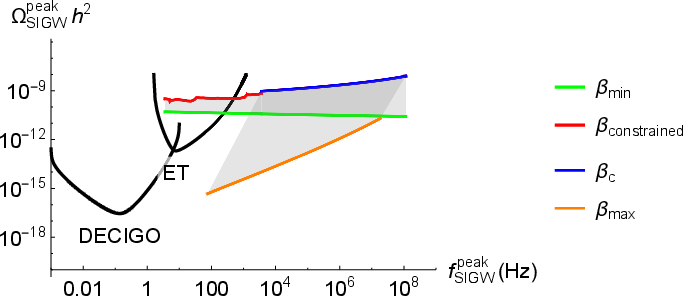}
    \caption{Our conjecture predicts null detection of SIGW that peaks inside the shaded gray regions, which corresponds to nonexistence of BHs with mass below $10^{15}\rm g$. See texts for detailed descriptions.}
    \label{fig:null test}
\end{figure}

We show in Fig. \ref{fig:null test} the shaded gray regions in which SIGW should not peak according to our conjecture. The sensitivity contours of upcoming GW detectors such as DECIGO and ET were also shown for references \cite{Schmitz:2020syl}. The considered mass range is $1\rm g\lesssim M_i\lesssim 10^{15}\rm g$ corresponding to peak frequency $1\text{ Hz}\lesssim f_\text{SIGW}^\text{peak}\lesssim 10^8\text{ Hz}$. The required lower bound on BH mass to not have information paradox (Eq. \ref{eq:M_BH}) is quite insensitive to the Universe's lifetime, thus probing null detection below $10^{15}\rm g$ is considered sufficient for a variety of crunching scenarios. At the other end, PBHs formed from inflationary perturbations must be heavier than 1 gram due to the observed upper bound on tensor-to-scalar ratio \cite{Loc:2025mzc}.

From Fig. \ref{fig:null test}, we have some comments:
\begin{itemize}
    \item The shaded region between the $\beta_\text{min}$ line and the $\beta_\text{constrained}$ line corresponds to $10^9\rm g\lesssim M_i\lesssim 10^{15}\rm g$. PBHs in this mass range evaporated after BBN and are significantly constrained by various Hawking-radiation-related observations \cite{Carr:2020gox}. Thus, the upper bound on the amplitude of SIGW depends on $\beta_\text{constrained}$ which is the upper bound on the PBH fraction obtained from experiments. We also noticed a recent study \cite{An:2026dfj} which sets even stronger constraints on PBH in the mass range between $10^{13}\rm g$ and $10^{14}\rm g$ from null detection of SIGW at  LIGO–Virgo–KAGRA, which further supports our conjecture.
    \item The shaded region between the $\beta_\text{min}$ line and the $\beta_c$ line corresponds to PBH mass in the range $1\rm g\lesssim M_i\lesssim 10^9\rm g$ and there is no PBH domination with $\beta<\beta_c$. In this case, apparently the upper bound on the amplitude of SIGW comes from $\beta_c$.
    \item The shaded region between the $\beta_\text{max}$ line and the $\beta_c$ line corresponds to PBH mass in the range $1\rm g\lesssim M_i\lesssim 10^9\rm g$ and there is PBH domination with $\beta>\beta_c$. In this case, more population and/or more mass means longer duration of PBH domination, which leads to more suppression of SIGW amplitude and more shift toward lower frequency. However, the existence of $\beta_\text{max}$ coming from the $\Delta N_\text{eff}$ constraint on isocurvature GW gives the maximum limit on the amount of amplitude suppression and frequency redshift. Also, because the SIGW amplitude is insensitive to $\beta$, the suppression effect of PBH domination significantly dominates over the enhanced GW amplitude of large $\beta$.
    \item  We note that this is the $\Omega_\text{SIGW}^\text{peak}-f_\text{SIGW}^\text{peak}$ plane. High-frequency tail of SIGW that peaks at lower frequency (corresponding to $M_i\gtrsim 10^{15}\rm g$) is sill allowed to exist in this region, but we expect it could barely touch the left edges. At the other end, it is impossible to have SIGW that peaks at frequency higher than $10^8\ \rm Hz$ due to lower bound on PBH mass ($M_i\gtrsim 1\rm g$) coming from the observed upper bound on tensor-to-scalar ratio, and thus low-frequency tail of SIGW cannot exist in this region.
\end{itemize}

\section{Discussion}\label{sec:discussion}

\subsection{Summary}
In the standard $\Lambda\rm CDM$ model of cosmology that stands for almost three decades, we don't have to even question whether or not the BH information paradox will ever happen because it's just a matter of time, and we have as much time as we need if the Universe just keeps expanding forever. However, we learned from DESI that DE is likely dynamical and that the Universe's lifetime may be finite. This dramatic shift of perspective leads us to question about the feasibility for the occurrence of BH information paradox during the Universe's lifespan. We proposed a conjecture called ``principle of no paradoxical Universe" in which the Universe has a configuration such that the ``anticipated" BH information paradox can never actually happen in reality. We estimated a lower bound on BH mass as a function of the Universe's lifetime to realize such scenario and some resulting predictions were made. Based on current data, the simplest crunching model predicts that Universe's lifetime is about 35 billion years, which implies nonexistence of BHs with mass below $\sim 10^{15}\rm g$ according to our conjecture. BHs formed from inflationary perturbation in the primordial Universe have a lower mass bound of order 1 gram to satisfy current constraint of tensor-to-scalar ratio. Therefore, the single most important prediction of our conjecture is null detection of SIGW with peak frequency in the range $\sim 1\ \rm Hz-10^8\ \rm Hz$ associated with PBH formation with mass $\sim 1\ \rm g-10^{15}\ \rm g$. We identified a region on the GW map in which SIGW should not peak. A portion of the parameter space could be tested in the upcoming ET detector, but new detectors operating in the $\sim100\ \rm Hz-10^8\ \rm Hz$ band are still needed to confirm our conjecture. If our conjecture is indeed correct, it will trigger many provocative thoughts on the fundamental nature of the Universe, some of which are discussed in the next subsection.

\subsection{Final thoughts}

Our conjecture could open up endless possibilities of follow-up research at the philosophical, mathematical, and physical levels. We will briefly list some interesting thoughts that came to our mind:

\begin{itemize}

\item While mutual contradictions between established physical laws lead to paradoxes that could signal new physics, our viewpoint is to think more deeply about these paradoxes and identify, under what conditions, if they are feasible in reality. In the case of BH information paradox, it is the timescale for the occurrence of paradox that matters. Thinking broadly, this simple observation hints at a much deeper question for the ontology of paradoxes: If the inconsistency among physical laws happens outside the regime of reality, what does it mean to say that we have a paradox? Should we consider ``foreseeable paradox" as a real thing to necessitate new physics, or should we consider singular point of pure logic as an imaginary vision that available natural resources can never afford?

\item An interesting aspect is that our bound is very close to (but slightly stronger than) the bound on PBH mass coming from the Trans-Planckian Censorship Conjecture (TCC). The TCC states that sub-Planckian modes should remain quantum in order to have a consistent theory of quantum gravity \cite{Bedroya:2019snp}, which leads to an upper bound on the energy scale of inflation $V^{1/4}<6\times 10^8\ \rm GeV$ \cite{Bedroya:2019tba}. By using $H_\text{inf}\simeq V^{1/2}/\sqrt{3}M_\text{pl}$ in the slow-roll approximation with $M_\text{pl}= 2.43\times 10^{18}\ \rm GeV$ is the reduced Planck mass, the corresponding upper bound on the Hubble rate during inflation is $H_\text{inf}<0.09\ \rm GeV$. In the simplest scenario of instant reheating, using $T_\text{reh}=(90M_\text{pl}^2H_\text{inf}^2/\pi^2g_{*,\rm reh})^{1/4}=(30V/\pi^2g_{*,\rm reh})^{1/4}$ leads to an upper bound on reheating temperature $T_\text{reh}<2.46\times 10^8\ \rm GeV$ where $g_{*,\rm reh}\sim 106.75$ has been used assuming Standard Model relativistic d.o.f. . By using Eq. \ref{eq:Mi vs Ti}, we get a lower bound on the PBH mass: $M_i>3\times 10^{14}\ \rm g$. Whereas our conjecture predicts $M_i>7.5\times 10^{14}\rm g$ for the simplest crunching model based on current data. Therefore, if all BHs existing in our Universe indeed respect the lower bound on mass proposed in our paper, they can have a primordial origin that is perfectly consistent with the TCC, have no paradoxes, and at the same time be a DM candidate ($M_i\sim10^{17}\rm g-10^{23}\rm g$).

\item If one takes into account other effects such as the quantum memory burden effect \cite{Dvali:2020wft,Dvali:2024hsb} or regular BH models \cite{Lan:2023cvz,Bambi:2023try}, the evaporation timescale of BH could be modified \cite{Loc:2024qbz,Loc:2025mzc}. Thus, the lower bound on initial  BH mass and the resulting physical predictions would be shifted accordingly. Here, we are working on the minimal assumption of singular Schwarzschild BH with standard semi-classical evaporation.

\item There is a class of related ideas known as ``bouncing cosmology" \cite{Novello:2008ra}. In this scenario, the Universe undergoes multiple expanding and contracting phases, and the current Universe is just a stage of that (potentially) endless process. It is an alternative solution for the horizon and flatness problems without the need of an inflationary period and there is no singularity like the Big Bang scenario. In this case, some fraction of BHs may still stay alive through different phases if the final size of the Universe after the contracting phase is still big enough \cite{Clifton:2017hvg}. In this case, a surviving BH that did not reach the Page time in the previous phase may eventually reach that timescale in the subsequent phase, so the information paradox is still there. There are two ways to think about this: 1) Our scenario is \textit{not} bouncing cosmology since all we suggest is that there will be a big crunch in the future in which the Universe recollapses to a singularity and obviously no BH could survive. Thus, this bouncing argument is irrelevant. 2) Even assuming bouncing cosmology, if the size of the Universe in the previous contracting phase is small enough, then BHs could have merged and created bigger BHs with extended lifetime. Thus, the Page time may still never be reached in the subsequent expanding phase. This second scenario is an interesting avenue for upcoming research.

\item There are typically two ways to think about our conjecture: 1) The Universe simply chose a configuration such that the BH information paradox does not occur in reality. There is no need to ask ``why" because it's just a brute fact. It is the way it is, just like we don't need to ask ``why is there something rather than nothing?". As long as the scenario is consistent with physical laws and tested by observations, that's all we care about. 2) If one wants to think about our conjecture from a dynamical perspective, one can speculate that perhaps there is a fundamental, yet-to-be-known correlation between BH's evolution and cosmic evolution. This mindset has a long tradition and motivated plenty of research directions such as the formation of regular BH or similar exotic singularity-free object with a de Sitter core, or cosmologically coupled BH, and so on. It would be interesting to rethink these decades-old ideas in light of our novel conjecture.

\end{itemize}

\section*{Acknowledgements}
This work was supported by the National Science and Technology Council, the Ministry of
Education (Higher Education Sprout Project NTU-114L104022-1), and the National Center
for Theoretical Sciences of Taiwan.

\appendix
\section{Dynamical dark energy}\label{sec:basics of dynamical DE}

We present in this appendix some essential aspects of dynamical DE that are most relevant for our present purpose. We then apply the framework to calculate the lifetime of the Universe in a quadratic Hilltop quintessence model.

\subsection{Field dynamics}
The simplest possible model for dynamical DE is the rolling of a minimally coupled scalar field $\phi$ down its potential $V(\phi)$. The Lagrangian is given by
\begin{equation}
    \mathcal{L}=\frac{1}{2}\nabla_\mu\phi\nabla^\mu\phi-V(\phi),
\end{equation}
where $\nabla_\mu$ is covariant derivative with respect to variable $x^\mu$. The $(+,-,-,-)$ signature is adopted here. The Klein-Gordon equation in an expanding Universe is then
\begin{equation}\label{eq:KG equation}
    \ddot{\phi}+3H\dot{\phi}+\frac{\partial V(\phi)}{\partial\phi}=0.
\end{equation}
The case of a static spacetime is recovered in the limit $H\rightarrow 0$. In an expanding Universe, the term $\propto H$ plays the role of friction to slow down the rolling of the field down the potential.

Sometimes it is also convenient to have the EOM expressed in terms of the scale factor instead of time:
\begin{equation}\label{eq: EOM_Hamiltonian}
        \begin{cases}
            \dot{\phi}=\pi_\phi\\
            \ddot{\phi}+3H\dot{\phi}+\frac{\partial V}{\partial\phi}=0
        \end{cases}
        \Rightarrow
        \begin{cases}
            \frac{d\phi}{da}=\frac{\pi_\phi}{aH}\\
            \frac{d\pi_\phi}{da}=-\frac{1}{aH}\frac{\partial V}{\partial\phi}-3\frac{\pi_\phi}{a}
        \end{cases}
\end{equation}
where $\pi_\phi$ is the momentum of the field and the time variable was converted into the scale factor variable by using $d/dt=\dot{a}\ d/da=aH\ d/da$. From the past to the present, the Hubble rate is positive and thus is well-defined as
\begin{equation}\label{eq: H_Quintessence}
    H=\sqrt{\frac{8\pi G}{3}\rho_\text{tot}}=\sqrt{\frac{8\pi G}{3}\left(\sum_i\rho_i+\rho_\text{DE}\right)}=\sqrt{\frac{8\pi G}{3}\left(\rho_\text{crit,0}\sum_i\frac{\Omega_{i,0}}{a^{3(1+w_i)}}+\frac{\pi_\phi^2}{2}+V(\phi)\right)},
\end{equation}
where we used the definition that $\Omega_{i,0}=\rho_{i,0}/\rho_\text{crit,0}$ with $\rho_{i,0}$ is the present-day energy density of the ith specie, the critical energy density today is $\rho_\text{crit,0}=3H_0^2/8\pi G$ with $H_0$ is the Hubble constant assuming a flat Universe, the fact that $\rho_i\propto a^{-3(1+w_i)}$ for constant equation of state $w_i$, the scale factor today is set to unity $a_0=1$, and $\rho_\text{DE}=\pi_\phi^2/2+V(\phi)$.

Conservation of energy of a fluid with energy density $\rho$ and pressure $p$ in an expanding FRW Universe gives
\begin{equation}
    \dot{\rho} + 3H(\rho + p) = 0.
\end{equation}
The equation of state (EOS) parameter is defined as $w(a)\equiv p(a)/\rho(a)$ which may be a function of the scale factor $a$ to take into account possible time-dependence. We then have
\begin{equation}\label{eq: fluid}
    \dot{\rho} + 3H(1+w(a))\rho = 0,
\end{equation}
or in terms of the scale factor variable as
\begin{equation}
\dot{a}\frac{d\rho}{da}
+ 3H(1+w(a))\rho = 0.
\end{equation}
\begin{equation}
  \Rightarrow  \frac{d\rho}{\rho}
= -3\frac{1+w(a)}{a}da.
\end{equation}
\begin{equation}
\Rightarrow\int_{\rho_0}^{\rho(a)} 
\frac{d\rho'}{\rho'}
=-3\int_{1}^{a} \frac{1+w(a')}{a'}\,da'.
\end{equation}
\begin{equation}
\Rightarrow
\ln\frac{\rho(a)}{\rho_0}
= -3\int_{1}^{a}\frac{1+w(a')}{a'}\,da'.
\end{equation}
\begin{equation}\label{eq: rho(a)}
   \Rightarrow \frac{\rho(a)}{\rho_0}
= \exp\left[-3\int_{1}^{a}\frac{1+w(a')}{a'}\,da'\right].
\end{equation}
Here, the present scale factor is set to unity $a_0=1$ and the present-day energy density is $\rho_0$. Instead of the scale factor, one can also convert this relation into the redshift language by a change of variable $1+z=1/a$. This is the most general equation to describe the evolution of energy density of fluid. If the EOS $w$ is constant, we recover the standard relation $\rho\propto a^{-3(1+w)}$. If $w$ is time-dependent, one can then use specific EOS parameterizations such as the CPL parameterization used by the DESI collaboration to demonstrate dynamical DE.

The EOS parameterization is useful for phenomenological studies. But if we have a theoretical model, it is better to solve the EOM of the dynamical field to get the actual EOS:
\begin{equation}\label{eq:w_phi}
    w_\phi=\frac{p_\phi}{\rho_\phi}=\frac{K_\phi-V_\phi}{K_\phi+V_\phi},
\end{equation}
where $K_\phi$ is the kinetic energy density and $V_\phi$ is the potential energy density of the field. A canonical homogeneous field has the kinetic term $K_\phi=\dot{\phi}^2/2$.

\subsection{Age of the Universe}
We should first emphasize that the ``age" of the Universe must be understood as the amount of cosmic time that has been elapsed from the Big Bang to the present, which is different from the ``lifetime" (or lifespan) of the Universe that is defined as the amount of cosmic time elapsed from Big Bang to Big Crunch. In other words, the lifetime of the Universe equals its age plus any remaining time it has.

As long as the Universe has been expanding since the Big Bang, which means that $H>0$ from the past to the present, we can write the Hubble rate as
\begin{equation}
    H=H_0\sqrt{\sum_i\frac{\Omega_{i,0}}{a^{3(1+w_i)}}+\Omega_\text{DE,0}\ \exp\left[-3\int_1^a\frac{1+w_\text{DE}(a')}{a'}da'\right]}.
\end{equation}
By noting that $H=\dot{a}/a$, the age of the Universe at the scale factor $a$ can be calculated as
\begin{equation}\label{eq:t_age}
    t_\text{age}(a)=\frac{1}{H_0}\int_0^a\frac{da'}{a'\sqrt{\sum_i\frac{\Omega_{i,0}}{a'^{3(1+w_i)}}+\Omega_\text{DE,0}\exp\left[-3\int_1^{a'}\frac{1+w_\text{DE}(a'')}{a''}da''\right]}}.
\end{equation}
In here and what follows, the notation $w_i$ indicates the constant EOS of any material that is \textit{not} DE, whereas the notation $w_\text{DE}$ indicates the EOS of DE that may or may not be constant. The present age of the Universe is $t_\text{age}(a_0=1)$.

\subsection{Lifetime of the Universe}
To calculate the lifetime of the Universe that may have a crunching phase in the future, we must be cautious that $H<0$ in the future. We thus need to look at the quadratic form of the Friedmann equation
\begin{equation}
    H^2=\frac{8\pi G}{3}\rho_\text{tot}=\frac{8\pi G}{3}\left(\sum_i\rho_i+\rho_\text{DE}\right).
\end{equation}
By taking derivative with respect to $t$, we get
\begin{equation}
    2H\dot{H}=\frac{8\pi G}{3}\left(\sum_i\dot{\rho_i}+\dot{\rho}_\text{DE} \right)=-8\pi GH\left(\sum_i(1+w_i)\rho_i+(1+w_\text{DE})\rho_\text{DE} \right)
\end{equation}

\begin{align}
       \Rightarrow \dot{H}&=-4\pi G\left(\sum_i\frac{\rho_\text{i,0}}{a^{3(1+w_i)}}(1+w_i)+\rho_\text{DE}(1+w_\text{DE})\right)
   \\
   &=-4\pi G\left(\sum_i\frac{\rho_\text{i,0}}{a^{3(1+w_i)}}(1+w_i)+\dot{\phi}^2\right)
\end{align}

\begin{equation}\label{eq: dimensionless H}
    \Rightarrow \mathbb{H}'\equiv\frac{\dot{H}}{H_0^2}=-\frac{3}{2}\left(\sum_i\frac{\Omega_{i,0}}{a^{3(1+w_i)}}(1+w_i)+\frac{8\pi G}{3}\phi'^2\right)
\end{equation}
where we used Eq. \ref{eq: fluid} in the first step and Eq. \ref{eq:w_phi} in the second step. We also defined the dimensionless Hubble rate $\mathbb{H}\equiv H/H_0$ and the prime denotes derivative with respect to the dimensionless time variable $H_0t$. The Klein-Gordon equation of the field in this dimensionless time variable framework is
\begin{equation}\label{eq:KG in H0t}
    \phi''+3\mathbb{H}\phi'+\frac{1}{H_0^2}\frac{\partial V(\phi)}{\partial\phi}=0.
\end{equation}
By solving numerically the above two coupled equations supplemented by the relation $\mathbb{H}=a'/a$, with the initial conditions are fixed at the present time, we can solve for $\{a,\mathbb{H},\phi\}$ as a function of $H_0t$ to determine the time at which the scale factor goes to zero, which is the lifetime of our Universe.

\subsection{Application to quadratic Hilltop quintessence model}
\begin{figure}[h!]
    \centering
    \includegraphics[width=0.49\linewidth]{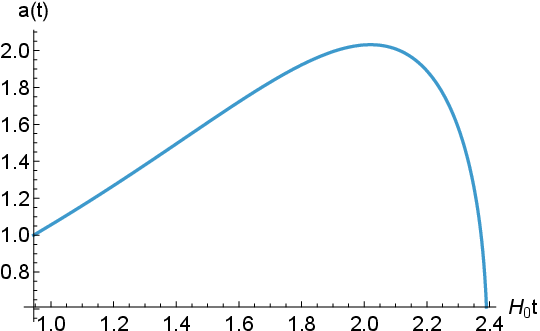}
    \includegraphics[width=0.49\linewidth]{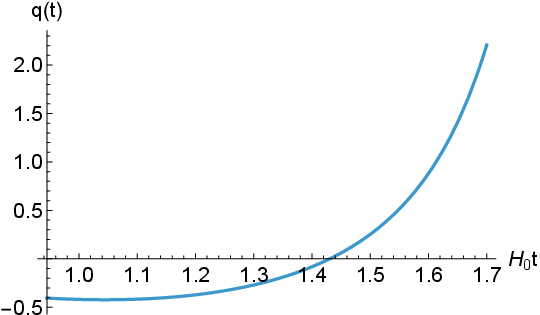}
    \includegraphics[width=0.49\linewidth]{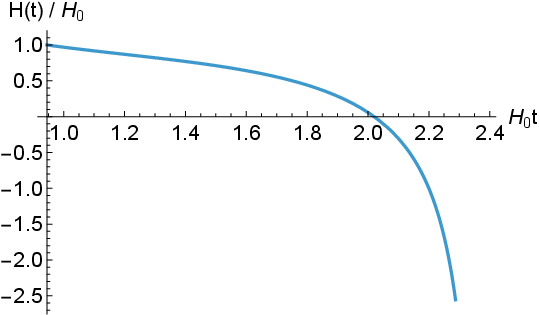}
    \includegraphics[width=0.49\linewidth]{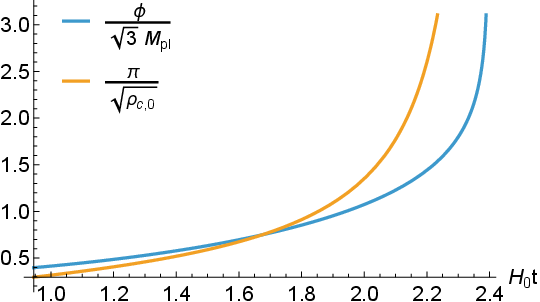}
    \includegraphics[width=0.49\linewidth]{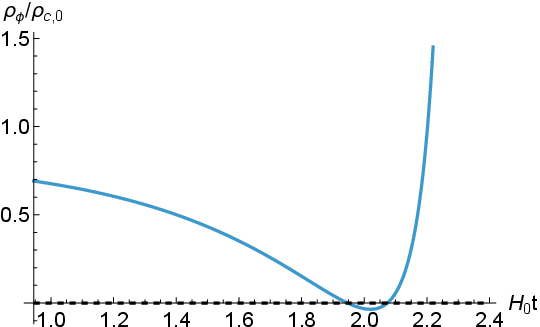}
    \caption{\textit{Upper left:} Evolution of the scale factor as a function of dimensionless time variable $H_0t$. \textit{Upper right:} Evolution of deceleration parameter. \textit{Middle left:} Evolution of the normalized Hubble rate. \textit{Middle right:} Evolutions of the normalized quintessence field and momentum. \textit{Bottom:} Evolution of the normalized quintessence energy density. The horizontal black dashed line indicates $\rho_\phi=0$.}
    \label{fig:a(t) and H(t)}
\end{figure}

After providing all the necessary tools, we now wish to work out an explicit example. One of the simplest quintessence models is the hilltop potential of the form
\begin{equation}
    V(\phi)=V_0+\frac{1}{2}V_2\phi^2,
\end{equation}
where $V_0>0$ and $V_2<0$. The rolling of the field toward the negative regime of the potential can realize a crunching scenario. By combining the datasets of Planck, DES, and DESI, the best-fit parameters of this model were found to be $V_0=3.26\times 10^{-47}\ \rm GeV^4$, $V_2=-6.51\times 10^{-84}\ \rm GeV^2$, $H_0=66.85\ \rm km/s/Mpc^{-1}$, and $\Omega_{m,0}=0.31$ assuming a flat Universe \cite{Cline:2025sbt} \footnote{In the dimensionless units of \cite{Cline:2025sbt}, the parameters are $V_0=0.9$, $V_2=-3.2$}. By solving the coupled equations of motion in Eq. \ref{eq: EOM_Hamiltonian} supplemented by the Hubble rate in Eq. \ref{eq: H_Quintessence}, we find the initial value of the field $\phi_i= 1.12\times 10^{18}\ \rm GeV$ can reproduce the observed DE abundance today $\Omega_\text{DE,0}=1-\Omega_{m,0}=0.69$. The initial conditions were set at $a_i=10^{-3}$ with the initial momentum $\pi_\phi=0$. Solving the system of Eqs. \ref{eq: EOM_Hamiltonian} and \ref{eq: H_Quintessence}, we can get the EOS $w_\phi(a)$ from the past to the present. Then by using Eq. \ref{eq:t_age}, we find that the age of the Universe is $t_\text{age}\approx 13.84\ \rm Gyr$, which is very close to standard $\Lambda \rm CDM$ result of $13.81\ \rm Gyr$ if using the central values of Planck data \cite{Planck:2018vyg} \footnote{This subtle difference comes from the fact that DE is weakening near the present according to DESI, so it takes slightly more time for the Universe to reach the present size.}. Thus, we can set the initial condition at $H_0t=0.945$ with $\mathbb{H}=1$ and $a=1$ to numerically solve the coupled Eqs. \ref{eq: dimensionless H} and \ref{eq:KG in H0t}
supplemented by $\mathbb{H}=a'/a$ to solve for $\{a(t),H(t), \phi(t)\}$. It is also useful to obtain the evolution of the deceleration parameter defined as $q\equiv -\ddot{a}a/\dot{a}^2=-(\dot{H}/H^2+1)=-(\mathbb{H}'/\mathbb{H}^2+1)$. The results are shown in Fig. \ref{fig:a(t) and H(t)}.

From Fig. \ref{fig:a(t) and H(t)}, we see in the upper left panel that the Universe reaches a maximum size of about 2 times its present size at $H_0t\approx 2.02$, corresponding to 29.57 Gyr since the Big Bang or 15.73 Gyr from now. It then begins to crunch to a singularity at $H_0t\approx 2.39$, corresponding to 34.99 Gyr after the Big Bang or 21.15 Gyr from now. Note that the crunching of the Universe is governed by a dynamical field and thus its lifetime is \textit{not} simply double of the time it reaches the maximum size, as opposed to the case of crunching phase driven by negative spatial curvature term or a negative cosmological constant. In fact, the fraction of total lifetime the Universe spending in the crunching phase is only $(2.39-2.02)/2.39\approx 15.48\%$. The upper right panel indicates that the Universe is still accelerating today with $q_0\approx -0.4$, but it will enter a decelerating phase when $q(t)$ flips its sign at $H_0t\approx 1.43$ corresponding to 20.94 Gyr since the Big Bang or 7.1 Gyr from now.

The middle left panel shows the associated Hubble rate where it changes the sign from positive to negative at $H_0t\sim 2.02$ when the Universe begins to crunch and hence $\dot{a}<0$. The middle right panel shows the evolution of quintessence field and its momentum normalized to the reduced Planck mass $M_\text{pl}$ and today's critical energy density $\rho_\text{c,0}$ for convenience. We see that the field evolves from a small initial value today to large value as the field rolls down the potential and its momentum increases. The bottom panel shows that $\rho_\phi$ drops below zero and flips its sign at $H_0t\approx 1.95$ as the field travels toward negative regime of the potential, which subsequently causes a crunching phase.

\section{A lower bound on population to claim existence of black holes} \label{appendix: beta_min}

We propose that a BH of initial mass $M_i$ can be said to exist in nature if at least one has ever existed within the observable Universe. There could be BHs with so small population that they or their Hawking radiation are existing outside the current observable Universe. However, because it is impossible for human beings to see anything outside of the observable Universe, it is impossible to confirm the existence of such BHs.

Assuming PBHs form in the standard radiation-dominated phase, the number of PBHs existing within the observable Universe at any given time is 
\begin{equation}
    N_\text{PBH}(t)=\frac{\beta\ \rho_\text{rad,i}}{M_i}\left(\frac{a_i}{a(t)}\right)^3\frac{4\pi}{3}R_\text{obs}^3(t),
\end{equation}
where $R_\text{obs}(t)$ is the physical radius of the observable Universe (i.e., particle horizon) at time $t$ and is given by
\begin{equation}
    R_\text{obs}(t)=a(t)\int_0^t\frac{dt'}{a(t')}.
\end{equation}
By noting that $\rho_\text{rad,i}=3H_i^2/8\pi G$ and $M_i=\gamma M_\text{H,i}=\gamma /2GH_i=\gamma\sqrt{45}m_\text{pl}^3/4\pi^{3/2}g_{\rho,i}^{1/2}T_i^2$, we arrive at
\begin{align}
    N_\text{PBH}(t)&=\frac{\beta\gamma^2m_\text{pl}^6}{8M_i^3}a_i^3\left(\int_0^t\frac{dt'}{a(t')}\right)^3\\
    &=\frac{\beta\gamma^2m_\text{pl}^6}{8M_i^3}\left(\frac{g_{s,0}T_0^3}{g_{s,i}T_i^3}\right)\left(\int_0^t\frac{dt'}{a(t')}\right)^3\\
    &=\frac{\beta\gamma^{1/2}m_\text{pl}^{3/2}g_{s,0}T_0^3\pi^{9/4}}{g_{\rho,i}^{1/4}45^{3/4}M_i^{3/2}}\left(\int_0^t\frac{dt'}{a(t')}\right)^3,
\end{align}
where $g_s$ is number of effective d.o.f. in entropy and at high temperature $g_\rho= g_s$, the subscript ``0" denotes the present values. We used the fact that entropy is conserved when going from the first line to the second line because apparently there is no PBH domination when we are looking for a minimum population. The condition to have at least one BH ever existed within the observable Universe is then
\begin{equation}
    N_\text{PBH}(t_\text{age})\geq 1,
\end{equation}
which gives
 \begin{equation}
    \beta\gtrsim 8\times10^{-83}\gamma^{-1/2}\left(\frac{g_{\rho,i}}{106.75}\right)^{1/4}\left(\frac{M_i}{\rm g}\right)^{3/2}.
\end{equation}
Here, $a(t)$ was obtained by numerically solving the Friedmann equation assuming the standard flat $\Lambda\rm CDM$ model with cosmological parameters $\Omega_{m,0}=0.32,\ \Omega_\text{rad,0}=8.6\times 10^{-5},\ \Omega_{\Lambda,0}=1-\Omega_{m,0}-\Omega_\text{rad,0},\ H_0=67\ \rm km/s/Mpc=2.17\times 10^{-18}s^{-1}$ \cite{Planck:2018vyg} and other numerical factors are $g_\text{s,0}=3.94,\ T_0=2.3\times 10^{-13}\rm GeV$. This lower bound applies to both evaporated and evaporating PBHs.  If PBHs evaporated outside of the observable Universe at the time of evaporation, their remnants (either stable compact object or Hawking radiation) could go inside the observable Universe at a later time (since particle horizon grows over time) and leave observable imprints. Therefore, we chose the upper bound of the integral to be evaluated at the current age of the Universe to make sure including any potential remnants that have ever existed inside the current maximum observable Universe, which is around 92 billion light years in diameter \footnote{In some literature, we may see people quote 93 billion light years. That is because they used $H_0=70\ \rm km/s/Mpc^{-1}$ inferred from local measurements.}.

\bibliographystyle{JHEP}
\bibliography{references}

\end{document}